# Exploiting synchrotron X-ray tomography for a novel insight into flax-fibre defects ultrastructure

[a]Delphine Quereilhac, [a]Lola Pinsard, [b]Elouan Guillou, [a]Marina Fazzini, [a]Emmanuel De Luycker, [c]Alain Bourmaud, [d]Marwa Abida, [e]Jonathan Perrin, [e]Timm Weitkamp, [a]Pierre Ouagne

[a]Univ. de Toulouse, Laboratoire Génie de Production, LGP, INP-ENIT, F-65016 Tarbes, France
[b]IPC Laval, Rue Léonard De Vinci, Changé, France
[c]Univ. Bretagne Sud, UMR CNRS 6027, IRDL, F-56100 Lorient, France
[d]Univ. de Lorraine, UMR-CNRS 7239, LEM3, F-57078 Metz cedex 03, France
[e]Synchrotron SOLEIL, F-91190 Saint-Aubin, France
*Corresponding author: delphine.quereilhac@enit.fr

**Abstract**
Flax fibres are valuable reinforcements for tomorrow's composites. However, defects called kink-bands, which mainly appear on fibres during the extraction and transformation phases, might affect their mechanical properties. Defects induced pores, within the kink-band are investigated in this work. They were morphologically explored using synchrotron phase-contrast X-ray microtomography, a technique that displays a sharp 3D representation of the pores. The study highlights the link between kink-bands and secondary cell wall ultrastructure. Pores are organised concentrically around the lumen, and their low thickness suggest that they are located at the interface between cellulose layers within S2 (G) layer. Moreover, the pores inclination with reference to the lumen axis follows the typical microfibrillar angle changes observed in the literature in the kink-band region. The volumes of the pores were measured, and a local increase in porosity was revealed in zones where defects are most severe along the fibre.

1. Introduction

Flax plants are a promising renewable alternative to synthetic fibres for reinforcing polymer composites. They present ecological advantages such as renewability and biodegradability while ensuring good mechanical properties in tension and vibration dumping (Le Duigou et al., 2011; Pil et al., 2016). However, these fibres are affected by defects in their internal structure called kink-bands or dislocations, which can lead to obvious areas of weakness and initiate cracking or debonding in the composites (Hughes, 2012; Le Duc et al., 2011).
There is no consensus, yet, as to the origin of these defects, especially regarding the exact moment they appear. Some researchers suspect that the defects are present at the growth stage of the flax straw, due to growing conditions and abiotic stress (Hänninen et al., 2012; Richely et al., 2022; Thygesen, 2011). Others believe that these factors are not sufficient, without major growth or lodging accidents, to generate such structural defects (Bourmaud et al., 2022). However, most of the literature agrees concerning the role of extraction methods on the occurrence of kink-bands (Kozlova et al., 2022). The interaction between tools and fibres during the mechanical transformation processes of flax extraction and preparation, including scutching, hackling, and drafting, can be aggressive on the flax straws and then induce local compressive loads on fibres, that increase the occurrence of dislocations (Grégoire et al., 2021; Thygesen et al., 2011; Zeng et al., 2015).

Flax fibres have a hierarchical organisation presented in Fig.1 It is composed of layers with different compositions and thicknesses: the primary cell wall (PCW) and the secondary cell wall (SCW). The primary cell wall (PCW) is the outer thin layer and does not have specific mechanical function. The secondary cell wall is itself composed of 3 sub-layers, referred to as S1, S2 (G) and S3 (Gn). S2 (G) is the main sub-layer, representing at least 90% of the total thickness and has strong longitudinal mechanical properties, due to high cellulose content and low fibrillar angle (Baley et al. 2018; Goudenhooft et al. 2019). The presence of kink-bands affects the structural organisation in all the layers in an elementary flax fibre, but especially the S2 (G) layer, located in the SCW, which is the thickest and is mainly responsible for the longitudinal mechanical properties of the fibres (Gorshkova et al., 2010; Sadrmanesh and Chen, 2019). S2 (G) is characterised by a high crystalline cellulose content, with cellulose micro fibrils embedded in a non-cellulosic matrix of polymers, mainly composed of hemicellulose and pectin. Cellulose micro fibrils are oriented at a microfibrillar angle (MFA) which is closely correlated with the mechanical properties of the fibres (Bourmaud et al., 2013). The appearance of kink-bands induces a high deviation in the MFA (from 5-8° up to 30°-40°), thereby leading to ultrastructure heterogeneities and pores (Melelli et al., 2021).

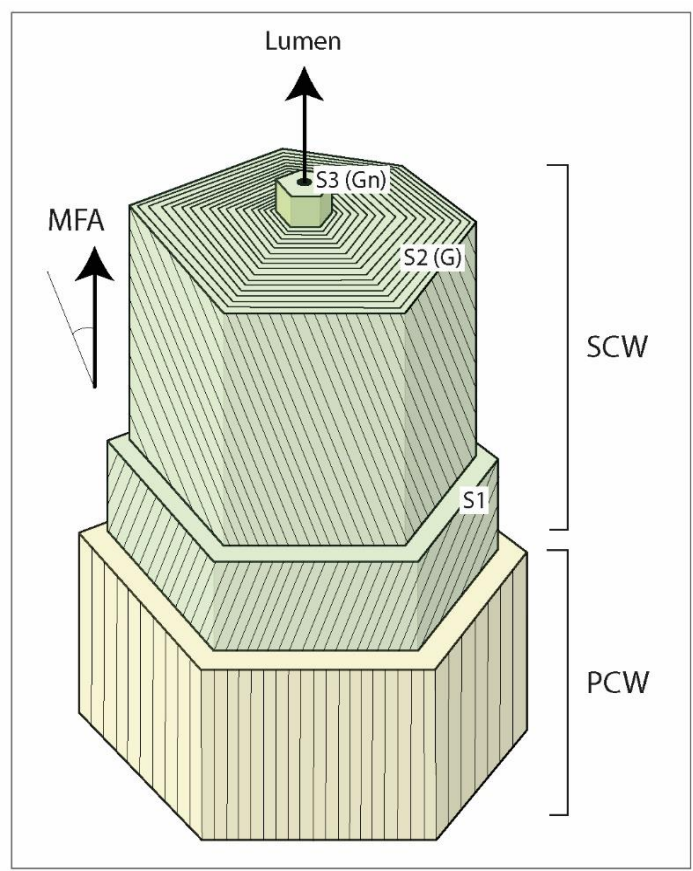

Figure 1 : Schematic representation of the multi-layer structure of a flax fibre with the primary cell wall (PCW) and secondary cell wall (SCW) composed of the S1, S2 (G) and S3 (Gn) layers

The lumen cavity, which runs the full length of the fibre, contributes naturally to the presence of porosity, but the occurrence of defects coincides with the sudden rise in the porosity rate, up to 6% of the cross-section surface area in kink-band regions (Bourmaud et al., 2022). The following research presented in this paper focuses on representing and understanding the organisation of the pores located within kink-bands. Many cutting-edge observation methods show representations of the defects at different scales but, unfortunately, mainly in 2D or multi plane representation. For instance, scanning electron microscopy (SEM) and atomic force microscopy (AFM) give an outer representation of the fibre, while second harmonic generation (SHG) and polarised light microscopy (PLM) give deeper insight into cellulose organisation (Melelli et al., 2020; Thygesen et al., 2006). The aim of this research work is to keep focusing on porosity organisation and volume and understand how their distribution can be linked with the secondary cell wall organisation. For this purpose, yet another method, namely, X-ray phase contrast microtomography, using synchrotron radiation, has been used on flax fibres and particularly on kink-bands areas. X-ray microtomography has already been used to observe plant cellulose walls and explain the damage mechanism (Beaugrand et al., 2017). Abbey et al. (Abbey et al., 2010) used synchrotron radiation X-ray tomography on flax fibres and evidenced the presence of pores inside the bulk of the cells but without clear and fine 3D visualization of kink-band volumes.

In this work, a novel three-dimensional description of the internal structure of individual flax fibres is given, with a sharp, close-up view of the kink-band pores. 3D representations of fibres were analysed to isolate the kink-band pores from the rest of the fibre components, giving new insight into the shape of the pores. A set of analyses was performed to describe the porosity distribution, shape and angles. A hypothesis about kink-band pores appearance in relation to the S2 (G) layer structure is also proposed.

## 2. Materials and methods

### 2.1. Flax plants and fibre extraction

Seeds of industrial Flax Bolchoï variety were sowed at the end of March 2020 by the company Van Robaeys Frères (Killem, France). Flax stems were pulled out in mid-July and then dew-retted in the field for 7 weeks. Then, after full retting, the fibres were successively extracted on a laboratory scutching and hackling line (Tarbes, France). The extraction line is composed of 3 separate modules, which perform the following:

1. Crushing, which consists in passing the straws between cylinders to break the woody core of the flax straw and release the fibres.
2. Beating, which removes wood and dust from the fibre's bundles, by passing them between two large rotating blades.
3. Hackling, which passes the flax fibre bundles through a series of increasingly smaller combs, to remove the last impurities, parallelise the fibres and separate the bundles into smaller ones.

The hackled fibres were then processed in the laboratory to produce a stretched sliver, which is a transformation step part of the spinning process. The resulting fibres thus underwent the classic processing steps found on an industrial scale.

### 2.2. X-ray microtomography experiments

Synchrotron-radiation X-ray phase contrast microtomography measurements were carried out at ANATOMIX beamline of Synchrotron SOLEIL (Weitkamp et al., 2017).

The instrument setup and configuration are described in detail in a previous publication (Bourmaud et al., 2022). In short, the fibres were mounted free-standing, protruding upward from a supporting needle to which they had been glued to avoid internal movement of the area under observation during the scan. The amount of glue was carefully chosen to avoid its penetration into the fibre pores. The measurement position, with a length of approximately 2 mm, was chosen slightly above the glue, to ensure mechanical stability while avoiding glue in the imaged zone. The measurements were then made using a polychromatic X-ray beam with a central photon energy around 12 keV. The detector had an effective pixel size of 0.325 µm and 2048 × 2048 pixels, resulting in a field of view of 0.65 × 0.65 mm2. The total exposure time for a full volume scan was around 4 minutes. Volume was reconstructed with voxels of the same size as the pixels (0.325 µm) using the standard data processing pipeline at the beamline, based on the software PyHST2 (ESRF, Grenoble, France) (Mirone et al., 2014). A Paganin filter (Paganin et al., 2002) was applied during the data reconstruction with a kernel length of 8 µm.

*2.3. 3D Image analysis*

The analysis of the reconstructed volume data, including segmentation, was performed with the Avizo version 2021.1 software (Thermo Fisher Scientific). After applying a non-local means filter to smooth and denoise the data, the materials were segmented by defining a threshold intensity value. Due to the different absorption contrasts between the background and the fibre cell wall tissues and the phase contrast in the signal, the difference between the pores and material was assessed visually.

First, all the pores were isolated from the rest of the internal fibre matter: a thresholding segmentation tool was used to isolate only the voids inside the fibres, which included lumen and kink-bands. They were represented by the lowest density and therefore the darkest grey level. Then the lumen was separated from the rest of the pores to isolate it from the kink-band porosity. In the defect-free regions, this work was simplified by the fact that the lumen is easily recognizable as it runs lengthwise through the fibre. In the defect zones, this separation was made using a very precise brush segmentation tool, to select only the lumen without including any kink-bands pores. Finally, the pores were isolated for volume measurements with an Avizo filter called "Separate objects". Each void volume has been assigned a colour to enable the observer to distinguish them.

## 3. Results and discussion

*3.1. Overall observations on porosity*

Bourmaud et al. 2022, explained that the kink-bands which are observed in flax fibres are the consequence of compressive loads applied during the fibre extraction from the stem and during further transformation of fibres prior to spinning. Fig.2 shows different representations of the fibres and their kink-bands. Fig.2c to 2e show tomographic slices of fibres along transverse planes, and Fig.2a and b and 2f to 2h show 3D views of the pore volumes including lumen and pores induced by defects. Fig.2a presents the reconstruction of the bulk of the flax bundle before segmentation. The presence of kink-band can already be predicted as nodes can be seen on the outer surface of the fibres (red arrows). Fig.2b highlights all the voids (lumen and pores induced by defects) obtained after segmentation. The 4 fibres analysed in the sample present a total of 12 distinctive kink-bands, for a cumulate length of fibre analysed of 2.66 mm. The main interest of working with a 3D representation is to decipher the complex distribution of the pores along the flax fibre. Kink-bands cannot be fully observed in a single plane: Fig.2c to e show a cross-section of an elementary fibre in three different positions along the Z axis, and Fig.2f shows the same defect in 3D. It is clear that a single slice is not enough to have a correct representation of the defect, as it moves all around the lumen when going forward along the Z axis. This is also valid for single slices in other orientations (orthogonal to the X and Y axes or oblique); full 3D data around the defect are therefore indispensable.

Fig.2f to h show three examples of defects that were observed when isolated from the rest of the fibre. The pores exhibit different morphologies and organisations (when they are not too crushed by the compressions, as this makes it difficult to accurately view the shape of the defects). The most frequently observed shapes are "planar" (Fig.2f), "X" (Fig.2g) and "helicoidal" (Fig.2h). These successive pores are induced by cell-wall deformation in the kink-band zone, but the specific conditions in which these various deformations develop are not clearly understood yet. However, the specific oblique shape of kink-band (diagonal with respect to the lumen axis) has already been observed in 2D with polarised light microscopy by Dinwoodie, and Keith and Cote (Dinwoodie, 1968; Keith and Cote, 1968) on wood fibres, but without highlighting the presence and the 3D shape of the pores. Cell-wall deformation was defined as "slip-line" to illustrate the deviation of cellulose micro-fibrils and microscopic compression crease. Nyholm et al., (2001) gave an extensive panel of the types of deformations encountered in wood and fibre cell-walls. According to them, these specific pore shapes can thus be linked with the external appearance of the fibre and the S1 layer deformation.

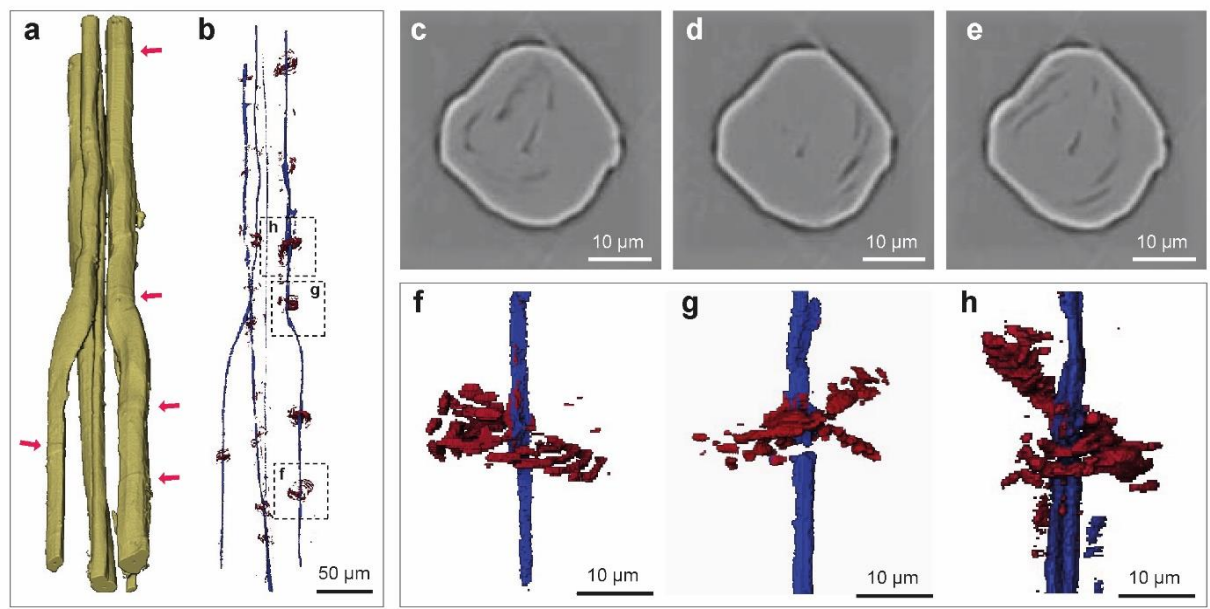

Figure 2: 3D visualization of the fibre, with red arrows pointing at nodes (a), voids within the fibre including lumen in blue and defect induced porosities in red (b), tomographic slices of fibres in the radial tangent plane where lumen and pores can be seen as dark shapes (c, d, e), 3D visualization of kink-band porosity in red around the lumen in blue (f, g, h)

*3.2. Evidencing the link between kink-band shaped porosity and S2 layer composition*

The S2 (G) layer cell wall can be up to 15 µm or 90% of the total cell wall area at maturity (Morvan et al., 2003), and is composed of approximately 75–90% cellulose, 15–20% hemicellulose (neutral polysaccharides) and 5–10% pectin (pectic polysaccharides) (His et al., 2001). As shown above, the organisation of the pores around the lumen can form different types of shapes, but they have in common a clear organisation in concentric layers sectors, comparable to "onion peels". Fig.3 shows results regarding the porosity organisation. Fig.3a to d presents pores viewed in a radial tangent plane, which most clearly shows the circle pattern. The pore ring separation is very clear in Fig.3a to c where pores are observed all around the lumen. By contrast, Fig.3d presents a kink-bands in which the "onion layers" are more difficult to see because of the severity of the defect and the coalescence of bigger pores, making it harder to identify successive layers. Two sets of analysis were performed on the images of the 12 distinct kink-bands (as seen on the fibre bundle in Fig.2b).

To begin with, the radius of the circular segments formed by the pore layer was measured to investigate its variation from one layer to the next. To this aim, a total of 7 zones in 4 different kink-bands were chosen, where at least three layers of pores were present and enough visible and suitable to fit within an associated circle. On a given sector of fibre (illustrated in Fig.3e), the circle was defined to fit as much as possible the pore shape. Once the circle was defined, its radius was recorded. The distance between the pores and the lumen was defined between the lumen centre of gravity (obtained using image analysis software ImageJ) and the midpoint of the fitting arc (intersection of the circle with the bisector of the fibre sector, illustrated Fig.3e). The evolution of the pores radius as a function of this previously defined distance was plot in Fig.3f for the different studied zones. Results show a linear increase in the pores radius close to identity, indicating that the kink-band pores form a set of concentric circles. This behaviour could be linked to the organisation of the cell wall growth, and especially to the S2 (G) layer organization and structure. This gelatinous zone is composed of successively forming cellulose layers during the growing stage with entrapped short galactan chains; at the Gn and immature stage, this layer is more porous and enriched in long galactan chains. At maturity, it develops into S2 (G) thanks to the galactosidase enzymatic activity, shortening the RG1 galactan chains and enabling stronger lateral interactions of cellulose micro-fibrils (Gorshkova et al., 2018). Nevertheless, the linear regressions do not seem to cut across the axis origin. This behaviour could be explained by the fact that flax fibres are not completely round (Haag and Müssig, 2016), and because the lumen void it-self is not always at the centre of the fibre.

Two hypotheses regarding the mechanism of tensile stress generation in the gelatinous layer could lead to pore formation: 1) longitudinal shrinkage of the cellulose network which could induce pressure on the outer cell-wall layers and create weaknesses at the interface of the cellulose layers (Mellerowicz and Gorshkova, 2012); 2) the entrapment of rhamnogalacturonan-I aggregates between the layers, limiting the lateral interaction of microfibrils (Gorshkova et al., 2018). Both hypotheses refer to zones of interaction between two layers in the gelatinous region. These mechanisms combined with the impact of flax extraction processes could lead to weaknesses and cause porosity to develop in the kink-bands. Moreover, it has been shown that the retting stage has an impact on cellulose organisation in flax fibres (8% increase in crystallinity with retting mainly due to the disappearance of amorphous polymer) and on its density (Bourmaud et al., 2019). These results give a structural explanation and justify the improvement in mechanical performance of the secondary wall in the flax fibres during the field retting process. The occurrence and severity of kink-bands could thus also depend on the retting level of the fibre.

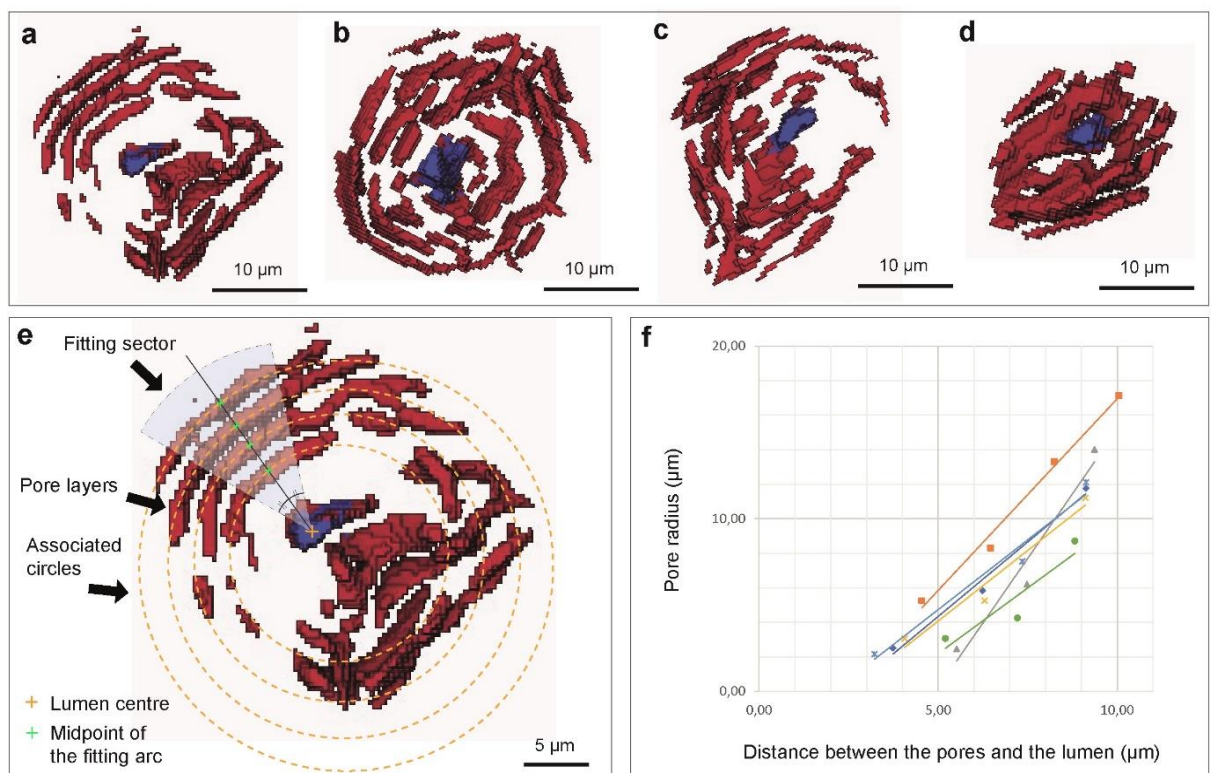

Figure 3: Defects viewed from the lumen axis, the lumen is coloured in blue, and the pore layers (in red) show a distinct onion layer organisation (a-d), detail of the measurement method for the pores radius (e), linear regression of the pores radius developing with the distance from the lumen axis (f)

Thickness measurements were then performed using ImageJ software on the kink-band slice images at several position in the pores to establish the size of successive pores and fibre layers (we assume these layers are made of gelatinous material which is preponderant in this area). Fig.4a shows two box plots representing the average thickness of the pores and the distance between them, for 120 measurement areas. Pore layer thickness (width of the pores in the radial direction) is very small, from 0.447 µm to 2.235 µm, with a median of 0.957 µm. At the fibre scale, the accumulation of pore layers in a kink-band region is responsible for the radial expansion of the fibre which leads to the nodes and elbow shape encountered on the external surface of the fibre in the defect zone. The uneven aspect of the fibre surface is due to the inside decohesion of successive cellulose layers. On the other hand, the distance between neighbouring pores (material thickness between adjacent pores) varies from 0.325 µm to 0.612 µm, with a median of 0.412 µm. The lower limit of these results corresponds to the resolution of the X-ray tomograph. It explains the low statistic repartition of the results (Fig.4a right plot) since pores smaller than the sensor resolution are filtered out from the data. Hock (Hock, 1942) focused on the cellulose layers in flax fibres, and showed that the individual layers varied in thickness from 0.1 to 0.2 µm, as represented in Fig.4b. It is assumed that the space between the pores is in the range of several layers of cellulose. Pores are spaced roughly at regular intervals which probably correspond to the interfaces of the different S2 (G) sublayers. These results support the hypothesis of an ordered interface porosity and not randomly distributed pores in the fibre bulk.

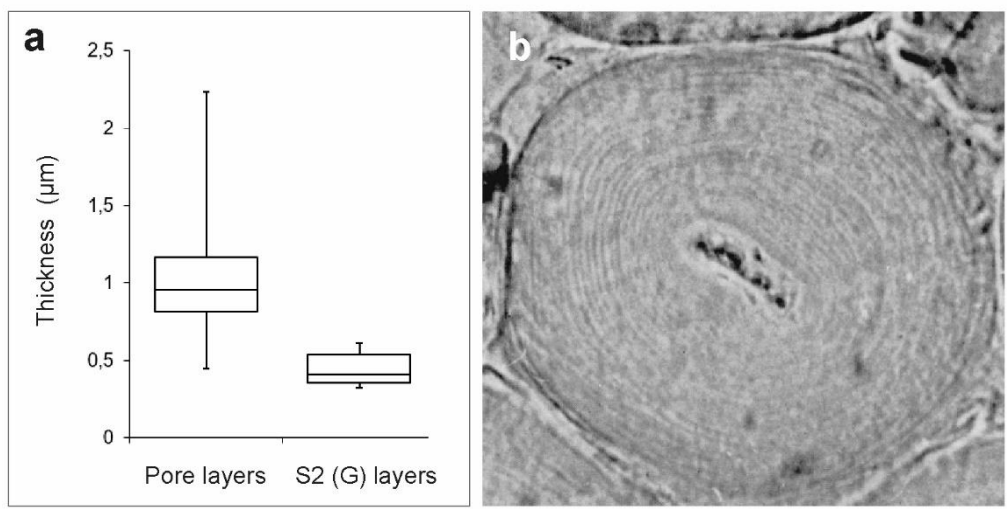

Figure 4: Box-plot presenting the pore layers thickness on the left, and the distance between two pores, corresponding to S2 (G) layers, on the right (a), cross-section of a flax fibre evidencing the successive cellulose layers (b) (Hock, 1942)

### 3.3. Pore volumes and kink-band severity

To continue the study of pores, their morphological parameters were measured. First, pores were individually segmented, as illustrated in Fig.5a. This segmentation was performed with a specific thresholding as some pores might "communicate" with others or with the lumen itself because of the kink-band severity resulting from the crushing of the fibre. This means that two successive pore layers will be considered as one if they touch each-other. The colours in Fig.5a were attributed randomly to represent the segmentation and separation of the volumes. After this step, it was possible to quantitatively determine the size and position of each pore. The box plot in Fig.5b presents a statistical analysis of the volumes of the 517 pores found in the fibres studied on the left and of a zone damaged by a kink-band shown in Fig.5a on the right. The results are very scattered, from 0.04 µm³ to 289 µm³, with a median of 0.34 µm³. Most of the pore volumes have very low values in this range, but they can reach up to 289 µm³ thereby locally increasing the pore volumes in the fibre bulk. It can be assumed that the very high values found in the left box plot in Fig.5b correspond to highly damaged areas. The layers were crushed against each other, touching and forming a single large pore on the image.

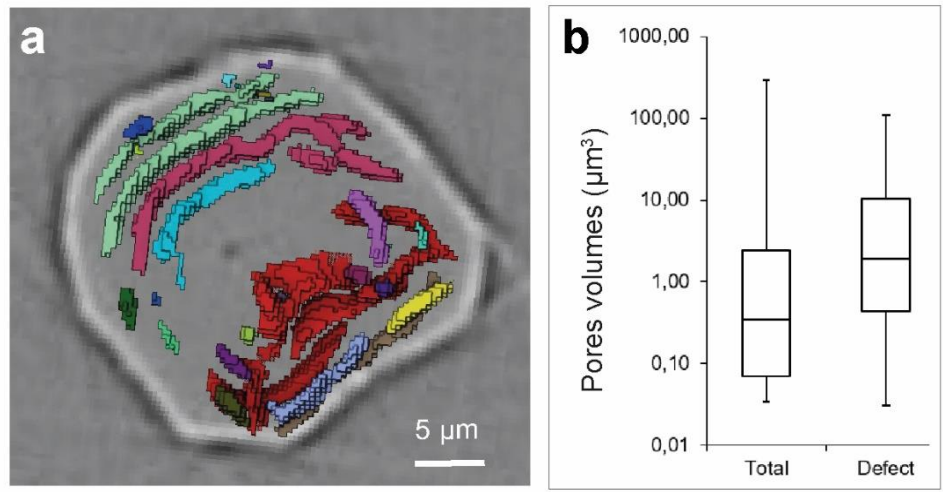

Figure 5: Focus on the defect already observed in Fig.3a, but here the pores were separated, colours help to see the distinction between pores (a), box-plot of the pore volumes generated by kink-bands, the left box represents the pore volume mean of all the zones studied, the right box represents the volume of pores of Fig.5a (b)

The difference between the left and right boxes in Fig.5b indicate that kink-band porosity might differ from one defect to the other, introducing a notion of severity which could be quantified by pore volume and/or number. This might also be explained by local differences in the fibre morphology that could impact the pore size, such as fibre diameter or extraction intensity. This notion of defect severity has already been explored in literature by comparing the size of defects against SEM or PLM, but the information obtained with these methods is limited (buckling length with SEM and bright zones with PLM). Counting kink-bands has been the subject of many studies (Baley, 2004; Kozlova et al., 2022; Thygesen, 2008; Thygesen and Hoffmeyer, 2005) because it may clarify the link with the mechanical behaviour of the fibre depending on the occurrence of kink-bands. For now, there is no clear conclusion concerning this link, as counting methods often do not take into account the severity of the defects. In fact, SEM gives only an external view, and PLM data depend on the orientation of the polariser and might not reveal all the damaged zones. Quantifying the pore volumes using tomography could be a reliable technique to link the kink-band size and its severity with the mechanical properties of a flax fibre. In this work, the link between defect porosity and fibre mechanical properties is not investigated as the fragments of fibre investigated by synchrotron tomography were too short to be tested mechanically, for example in traction tests. Moreover, synchrotron radiation might damage the fibres if the dose used for the measurements is too high.

### 3.4. Orientation of the pore layers and MFA

Exploration of the pore morphology in the tomography data indicates that the cross-sections of the pores in the longitudinal radial plane are elliptical with a major axis slightly inclined with respect to the lumen axis. This characteristic can be measured on 2D slices (Fig.6 a and b). The misalignment of pores vs. lumen axis was measured around seven distinct defect zones and it appears that the pore layers are oriented in a similar direction, forming an angle between 22 and 51° (Fig.5c) with the axis of the fibre. MFA values generally range between 0° and 10° in defect-free areas, and stronger deviations exist in kink-band regions.

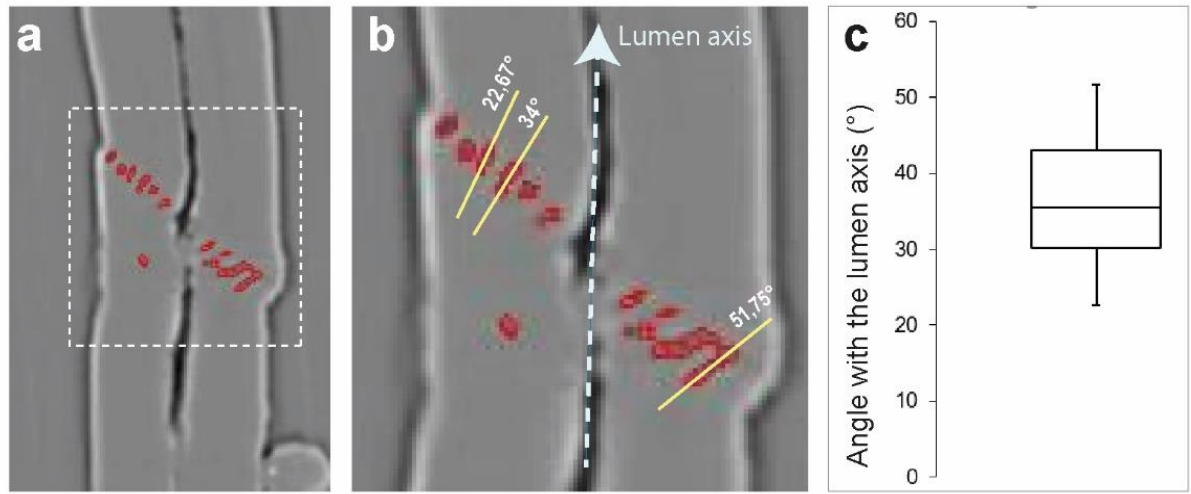

Figure 6: Tomographic slice images of fibres along longitudinal radial planes, where pores are highlighted (a), measurements of the angle of pores with the lumen axis (b), boxplots of 13 measured angles in zone where the misalignment is the most clearly visible (c)

While a hypothesis based on these results remains to be defined, it is noteworthy that the values obtained are very close to the misalignment of the MFA reported in literature for defect zones *(Melelli et al., 2021; Thygesen and Gierlinger, 2013)*. If we consider the previous proposition regarding the link between pore layer and S2 (G) cellulose layers, it appears logical that the deviated cellulose chains lead the pores in the same direction. This inclination of pores in the defects reinforces the idea about the link between porosity and the S2 (G) layer ultrastructure; the interfacial pores follow the cell wall layers when they are deviated during the kink-band occurrence.

**Conclusion**

In the present work, synchrotron phase-contrast microtomography was used to study the structural properties of flax fibre kink-band regions. As expected, this method complements other imaging techniques by yielding a sharp and detailed 3D representation of the pores. Thanks to this novel insight, previously unknown parameters regarding the pores were calculated, such as pore volumes, width, cell-wall layer thickness, and pore orientation.

The main findings of this research work are as follows:

- The shapes of the pores often follow defined patterns such as plane and X, as already seen in 2D with polarised light microscopy. The 3D representation of the pores offers the possibility to distinguish the full shape of the pore rings, and to find out their helicoidal organisation.
- The concentric layer structure and the inclination of the pores with the lumen axis (22°-51°) were established and it globally corresponds to the inclination of cellulose microfibrils.
- The thickness of S2 (G) sub-layers between pores is around 0.4 µm and could correspond to one or more layers of cellulose structure that form around the lumen during the cell wall thickening, according to the literature.
- Pore volumes generally remain under 4 µm$^3$ but were found to reach up to 289 µm$^3$, thereby inducing a considerable local increase in fibre size in the defect region.

In conclusion, these findings corroborate the idea that pore rings appear at cellulose layer interfaces in the gelatinous bulk. They are due to the decohesion of cellulose layers and misalignment of the MFA in the secondary cell wall. This work adds a new perspective in the techniques used to investigate kink-bands. The next step would be to go even further in these observations and increase spatial resolution, for example by using nanotomography (Scheel et al., 2021) to obtain an even more detailed view within the flax fibres to elucidate the relationship between porosity and S2 cell-wall layer construction. It would also be beneficial to use pore identification to link the presence of defects with the mechanical behaviour of the fibre. Finally, a step forward would be to use this information to go back to the exact occurrence of kink-bands at the industrial level, and to understand how the local increase in porosity impacts the mechanical behaviour of the fibre. The ultimate goal would be to minimise the presence and severity of defect zones in flax fibres in order to widen the panel of composite applications in which flax could be a substitute for synthetic fibres.

**Acknowledgements**

The authors wish to acknowledge the funding provided by the French National Research Agency (ANR) through the FLOEME project ANR-21-CE10-0008. ANATOMIX is an Equipment of Excellence (EQUIPEX) funded by the Investments for the Future programme of the French National Research Agency (ANR), project NanoimagesX, grant no. ANR-11-EQPX-0031. Access to Anatomix was provided through SOLEIL beamtime proposal #20201291. Elouan Guillou gratefully acknowledge IPC company, Laval Agglo and Région pays de Loire for the funding of his thesis.